\documentclass[structabstract]{aa}  
\usepackage{natbib}

\usepackage[]{hyperref}
\usepackage{amsmath}
\hypersetup{colorlinks=true,citecolor=blue}
\usepackage{graphicx}
\usepackage{draftcopy}

\usepackage{url}
\usepackage{amsmath}
\usepackage{amssymb}
\usepackage{amsfonts}
\usepackage[ruled]{algorithm2e}
\usepackage{graphicx,epsfig,graphics}
\usepackage{aas_macros}
\usepackage{graphics}
\usepackage{algorithmic}

\usepackage{amsmath}
\usepackage{graphicx}
\usepackage{subfigure}

\usepackage{txfonts}
\begin{document}

   \title{Spherical 3D Isotropic Wavelets}

   \author{Fran\c{c}ois Lanusse \inst{1}          
		  \and
		  Anais Rassat \inst{2,1}  
		 \and
		 Jean-Luc Starck \inst{1}  		  		  
   %       \fnmsep\thanks{Just to show the usage
   %       of the elements in the author field}
          }

   \institute{Laboratoire AIM, UMR CEA-CNRS-Paris 7, Irfu, SEDI-SAP, Service d'Astrophysique, CEA Saclay,
F-91191 GIF-Sur-YVETTE CEDEX, France.
 \and
      Laboratoire d'Astrophysique, Ecole Polytechnique F\'ed\'erale de Lausanne (EPFL), Observatoire de Sauverny, CH-1290, Versoix, Switzerland.}

\date{\today}

%\offprints{jstarck@cea.fr}

   \abstract{Future cosmological surveys will provide 3D large scale structure maps with large sky coverage, for which a 3D Spherical Fourier-Bessel (SFB) analysis in spherical coordinates is natural. Wavelets are particularly well-suited to the analysis and denoising of cosmological data, but a spherical 3D isotropic wavelet transform does not currently exist to analyse spherical 3D data.}{The aim of this paper is to present a new formalism for a spherical 3D isotropic wavelet, i.e. one based on the SFB decomposition of a 3D field and accompany the formalism with a public code to perform wavelet transforms.}{We describe a new 3D isotropic spherical wavelet decomposition based on the undecimated wavelet transform (UWT) described in Starck et al. 2006. We also present a new fast Discrete Spherical Fourier-Bessel Transform (DSFBT) based on both a discrete Bessel Transform and the HEALPIX angular pixelisation scheme. We test the 3D wavelet transform and as a toy-application, apply a denoising algorithm in wavelet space to the Virgo large box cosmological simulations and find we can successfully remove noise without much loss to the large scale structure.}{We have described a new spherical 3D isotropic wavelet transform, ideally suited to analyse and denoise future 3D spherical cosmological surveys, which uses a novel Discrete Spherical Fourier-Bessel Transform. We illustrate its potential use for denoising using a toy model. All the algorithms presented in this paper are available for download as a public code called {\tt MRS3D} at \url{http://jstarck.free.fr/mrs3d.html}}{}


\maketitle

\section{Introduction}

\subsection*{Challenges in Modern Cosmology}

The wealth of cosmological data in the last few decades \citep{wmap7,Schrabback:2010,Percival:2007} has led to the establishment of a standard model of cosmology, which describes the Universe as composed today of approximately 4\% baryons, 22\% dark matter and 74\% dark energy. The main challenges in modern cosmology are to understand the nature of both dark energy and dark matter, as well as the initial conditions of the Universe \citep{DETF,WGFC}. A thorough understanding of these three topics may lead to a revision of Einstein's theory of General Relativity and our view of the early Universe.

New surveys are planned who aim to answer these important questions e.g. Planck for the CMB \citep{Planck}, DES \citep[Dark Energy Survey,][]{DES:2005}, BOSS \citep[Baryon Oscillation Spectroscopic Survey,][]{Schlegel:2007}, LSST \citep[Large Synoptic Survey Telescope,][]{LSST} and Euclid \citep{Euclid:2011} for weak lensing and the study of large scale structure with galaxy surveys. 

The challenge with these upcoming large data-sets is to extract the cosmological information in the most suitable manner in order to test the cosmological paradigm. Depending on the signal one wishes to extract, and/or survey geometry, different bases may be more or less suitable (e.g., Fourier, Spherical Harmonic, Configuration or Wavelet Space). Moreover, future surveys may be in 2D (e.g. Planck) or in 3D (e.g. galaxy or weak lensing surveys). Where 3D data is available, a tomographic analysis is possible (also known as 2D 1/2), or a full 3D analysis can be done. For data in spherical coordinates, this corresponds to a Spherical Fourier-Bessel (SFB) decomposition \citep{Heavens:1995,Fisher:1995,weak3d,Erdogdu:2005wi,Erdogdu:2006dv,Leistedt2011,Rassat:2011bao}.

\subsection*{Wavelet Transform on the Sphere}

Wavelets are particularly well suited to the analysis of cosmological data \citep{aw:martinez93,starck:sta05_2}, since cosmological data can often be sparsely represented in wavelet space.  2D Wavelets have been used  in many astrophysical studies \citep{starck:book06} for a broad range of applications such as denoising, deconvolution, detection, etc.  
CMB studies have motivated the development of 2D spherical wavelet decompositions.  Continuous  wavelet transforms on the sphere~\citep{wave:antoine99,wave:tenerio99,wave:cayon01,wave:holschneider96} have been proposed, mainly for non Gaussianity studies.
In \citet{starck:sta05_2},  an invertible isotropic undecimated wavelet transform (UWT) on the sphere based on 
spherical harmonics was described, that can be also used for other applications 
such as  deconvolution, component separation \citep{starck:yassir05,bobin-gmca-cmb,delabrouille08}, inpainting \citep{inpainting:abrial06,starck:abrial08}, or Poisson denoising \citep{schmitt2010}.
A similar wavelet construction has been published  in \citep{marinucci08,fay08a,fay08} using so-called ``needlet filters", and in 
 \citet{wiaux08},  an algorithm was proposed which allows us to reconstruct an image from its steerable wavelet transform.
 Other multiscale  transforms on the sphere such as ridgelets and curvelets have been developed  \citep{starck:sta05_2}, and are well adapted to detect anisotropic features. Other multiscale transforms on the sphere, such as ridgelets and curvelets, have been developed \citep{starck:sta05_2} and are well adapted to detect anisotropic features. An extension of this UWT has also been developed for polarised CMB data  in \citet{starck:pola09}.

In this paper, we describe a new 3D isotropic spherical wavelet decomposition, which is reversible, and could therefore be useful for many different applications. It is based on the UWT proposed by \cite{starck:sta05_2} and extended into 3D. The 3D-UWT proposed here can be used to analyse 3D data in spherical coordinates, such as a 3D galaxy or weak lensing survey with large (but not necessarily full) sky coverage.

\section{Spherical 3D Filtering using the Spherical Fourier-Bessel Transform}

	\subsection{Spherical Fourier-Bessel Transform}
	\label{section:SFBT}
	
	The SFB transform of a square integrable scalar field $f(r,\theta,\phi)$ can be defined as:
\begin{equation}
	\hat{f}_{l m}(k) = \sqrt{\frac{2}{\pi}} \iiint f(r,\theta, \phi) j_{l}(k r)   \overline{Y_l^m(\theta,\phi)} r^2 \sin(\theta) dr d\theta d\phi,
\label{SFBT}
\end{equation}
where $Y_l^m$ are spherical harmonics and $j_{l}$ are spherical Bessel functions and $\overline{Y}$ represents the complex conjugate of $Y$. Note Equation \ref{SFBT} uses the same convention as \cite{Heavens:1995} which differs slightly from that of \cite{weak3d}, \cite{Leistedt2011} and \cite{Rassat:2011bao}, as explained below. This expression allows the expansion of a 3D field provided in spherical coordinates onto a set of orthogonal functions:
\begin{equation}
\Psi_{l m k}  (r, \theta, \phi) = \sqrt{\frac{2}{\pi}}j_{l}(k r) Y_l^m(\theta,\phi)
\end{equation}

From the orthogonality of the $\Psi_{l m k}$'s, the original field $f(r, \theta, \phi)$ can be reconstructed from its SFB transform $\hat{f}_{l m}(k)$ using the following inversion formula:
\begin{equation}
f(r, \theta, \phi) = \sqrt{\frac{2}{\pi}} \sum\limits_{l=0}^{\infty} \sum\limits_{m=-l}^{l} \int \hat{f}_{l m}(k) k^2 j_l(k r) dk Y_{l m}(\theta,\phi).
\label{Inv_FBD}
\end{equation}

From this definition, the SFB transform can also be regarded as the commutative composition of two different transforms: a Spherical Harmonics Transform (SHT) for the angular dimension and a Spherical Bessel Transform (SBT) for the radial dimension. In this work, the following convention is adopted to define the SHT of a function $f(\theta,\phi)$ defined on the sphere:
\begin{subequations}
\begin{align}
	f_{l m} & =  \int_0^{2\pi} \int_0^{\pi} \overline{Y_l^m(\theta,\phi)} f(\theta, \phi) \sin(\theta) d\theta d\phi  \label{SHT},\\
f(\theta,\phi) & =  \sum\limits_{l=0}^{\infty} \sum\limits_{m = -l}^l f_{l m} Y_l^m(\theta,\phi) \label{Inv_SHT}.
\end{align}
\end{subequations}

Our choice for $\Psi_{l m k}$ allows us to give a symmetrical expression for the SBT and its inverse:
\begin{subequations}
\begin{align}
\hat{f}_l(k) &  = \sqrt{\frac{2}{\pi}} \int f(r) j_l(k r) r^2 dr \label{SBT},\\
f(r) & =  \sqrt{\frac{2}{\pi}} \int \hat{f}_l(k) j_l(k r) k^2 dk \label{Inv_SBT}.
\end{align}\label{Def_SBT}
\end{subequations}
Although this symmetrical formulation for the SBT may differ from other works \citep[e.g., ][]{weak3d, Leistedt2011,Rassat:2011bao}, it will prove very convenient, especially to obtain a discretised transform (see Section \ref{sec:DSFBT}).

\subsection{Frequency filtering using the SFB Transform}

Spherical 3D filtering can be defined as the 3D convolution product of a 3D field in spherical coordinates with a 3D filter also provided in spherical coordinates. Using the relations presented in \ref{appendix:SFBT:Fourier} and \ref{appendix:SFBT:convolution}, it is possible to express such a product in terms of SFB coefficients and to relate those coefficients to regular Fourier coefficients.

To illustrate Spherical 3D filtering on a simple case, let us consider a simple isotropic low-pass filter $u$ whose 3D Fourier transform is $U(k,\theta_k, \phi_k)$. The 3D Fourier transform of such a filter has a spherical symmetry and takes a simple form in spherical coordinates. Indeed, $U(k,\theta_k,\phi_k)$ is a function only of $k$ because of its symmetry, therefore $U(k,\theta_k,\phi_k) = U(k)$. Furthermore, since $U(k,\theta_k,\phi_k)$ is constant on every sphere centred on the origin in Fourier space, its SHT for a given $k$ verifies $\forall (l,m) \neq (0,0), \quad U_{l m}(k) = 0$. As a result, the SFB transform of $u$ has the following simple expression:

\begin{equation}
	\hat{u}_{l m}(k) = \begin{cases}
	\frac{U(k)}{Y_0^0} & \mbox{if $l = m = 0,$} \\
	0				   & \mbox{otherwise.}
	\end{cases}
\end{equation}

Let us now consider a 3D field $f$ defined in spherical coordinates. The filtered field $v$ resulting from applying $u$ to $f$ is obtained using Eq. (\ref{Spherical_Convolution}) simplified by the properties of $g$:

\begin{eqnarray}
	\hat{v}_{l m}(k) & = & \widehat{(f \ast u)}_{l m}(k) ,\nonumber \\
	 & = & (i)^l \sqrt{(2 \pi)^3} \sum\limits_{l'=0}^{\infty} \sum\limits_{m' = -l'}^{l'} (-i)^{l'} \hat{f}_{l' m'}(k)\nonumber \\
	& & \times \quad \sum\limits_{l''= | l - l' |}^{ l + l'} c^{l''}(l,m,l',m') (-i)^{l''} \hat{u}_{l'' m-m'}(k) \delta_{l'' 0} \delta_{m'' 0}, \nonumber \\
								  & = & (-i^2)^l \sqrt{(2 \pi)^3}  \hat{f}_{l m}(k) c^{0}(l,m,l,m) (-i)^{0} \hat{u}_{0 0}(k).
\end{eqnarray}
Knowing that $c^{0}(l,m,l,m) = 1/\sqrt{4 \pi}$, the following expression is finally obtained:
\begin{equation}
	\hat{v}_{l m}(k) = \sqrt{2} \pi \hat{u}_{0 0}(k) \hat{f}_{l m}(k) .
	\label{Filtering_Spherical_Symmetry}
\end{equation}

In the special case of a 3D isotropic filter, frequency filtering is therefore easily obtained using the SFB transform and the filtered coefficients are simply the original coefficients multiplied by a function of $k$.

Although this filter seems overly simplistic, it will be shown in the following sections that such a low-pass filter is at the heart of the Isotropic Undecimated Spherical 3D Wavelet transform which makes direct use of Eq. (\ref{Filtering_Spherical_Symmetry}).

\section{Isotropic Undecimated Spherical 3D Wavelet Transform}\label{sec:wavelet}
	
An Isotropic Undecimated Spherical Wavelet transform defined on the sphere and based on the SHT was proposed in \citet{starck:sta05_2}.
Now the aim is to transpose the ideas behind this transform to the case of data in 3D spherical coordinates. Indeed, the isotropic wavelet transform can be defined using only an isotropic low-pass filter. In the last section the necessary relations to apply such a filter have been obtained and the Isotropic Undecimated Spherical 3D Wavelet transform can now be defined.

\subsection{Wavelet decomposition}

Using the formalism introduced in the previous section, a Wavelet transform can be defined with the SFB transform. This isotropic transform is based on a scaling function $\phi^{k_c}(r, \theta_r, \phi_r)$ with cut-off frequency $k_c$ and spherical symmetry. The symmetry of this function is preserved in the Fourier space and therefore, its SFB transform verifies $\hat{\phi}^{k_c}_{l m}(k) = 0$ whenever $(l,m) \neq (0,0)$. Furthermore, due to its cut-off frequency, the scaling function verifies $\hat{\phi}^{k_c}_{0 0}(k) = 0$ for all $k \geq k_c$. In other terms, the scaling function verifies:
\begin{equation}
	\Phi^{k_c} (r,\theta_r,\phi_r) = \Phi^{k_c} (r) = \sqrt{\frac{2}{\pi}} \int_0^{k_c} \hat{\Phi}^{k_c}_{0 0}(k) k^2 j_{0}(k r)dk Y_0^0 (\theta_r,\phi_r).
\end{equation}
Using relation (\ref{Filtering_Spherical_Symmetry}) the convolution of the original data $f(r,\theta,\phi)$ with $\Phi^{k_c}$ becomes very simple:
\begin{equation}
\hat{c}^0_{l m}(k) = \widehat{\left[\Phi_{k_c} \ast f\right]}_{l m}(k) = \sqrt{2} \pi \hat{\Phi}^{k_c}_{0 0}(k) \hat{f}_{l m}(k).
\end{equation}

Using this scaling function, it is possible to define a sequence of smoother approximations $c^j(r,\theta_r,\phi_r)$ of a function $f(r,\theta_r,\phi_r)$ on a dyadic resolution scale. Let $\Phi^{2^{-j}k_c}$ be a rescaled version of $\Phi^{k_c}$ with cut-off frequency $2^{-j}k_c$. Then $c^j(r,\theta_r,\phi_r)$ is obtained by convolving $f(r,\theta_r,\phi_r)$ with $\Phi^{2^{-j}k_c}$ :
\begin{eqnarray}
c^0 & = & \Phi^{k_c} \ast f ,\nonumber \\
c^1 & = & \Phi^{2^{-1} k_c} \ast f, \nonumber \\
	&\cdots & \nonumber \\
c^j & = & \Phi^{2^{-j}k_c} \ast f.
\end{eqnarray}
Applying the SFB transform to the last relation yields:
\begin{equation}
\hat{c}^j_{l m}(k) = \sqrt{2} \pi \hat{\Phi}^{2^{-j} k_c}_{0 0}(k) \hat{f}_{l m}(k).
\end{equation}
This leads to the following recurrence formula :
\begin{equation}
\forall k < \frac{k_c}{2^{j}}, \quad \hat{c}^{j+1}_{l m}(k) = \frac{ \hat{\Phi}^{2^{-(j+1)} k_c}_{0 0}(k)}{\hat{\Phi}^{2^{-j}k_c}_{0 0}(k) }\hat{c}^j_{l m}(k) .
\label{RecSmooth}
\end{equation}

Just like with the \textit{\`a trous} algorithm by \cite{starck:book10}, the wavelet coefficients $\{w^{j}\}$ can now be defined as the difference between two consecutive resolutions:
\begin{equation}
w^{j+1}(r,\theta_r,\phi_r) = c^j(r,\theta,\phi) - c^{j+1}(r,\theta,\phi).
\end{equation}

This choice for the wavelet coefficients is equivalent to the following definition for the wavelet function $\Psi^{k_c }$ :
\begin{equation}
\hat{\Psi}^{2^{-j} k_c }_{l m} (k) =  \hat{\Phi}^{2^{-(j-1)} k_c}_{l m}(k) -  \hat{\Phi}^{2^{-j} k_c}_{l m}(k),
\end{equation}
so that :
\begin{eqnarray}
w^0 & = & \Psi^{k_c} \ast f ,\nonumber \\
w^1 & = & \Psi^{2^{-1} k_c} \ast f, \nonumber \\
	&\cdots & \nonumber \\
w^j & = & \Psi^{2^{-j}k_c} \ast f.
\end{eqnarray}

By applying the SFB transform to the definition of the wavelet coefficients and using the recurrence formula verified by the $c^j$s yields:
\begin{equation}
\forall k < \frac{k_c}{2^{j}}, \quad \hat{w}^{j+1}_{l m}(k) = \left( 1 -  \frac{ \hat{\Phi}^{2^{-(j+1)} k_c}_{0 0}(k)}{\hat{\Phi}^{2^{-j}k_c}_{0 0}(k) } \right) \hat{c}^j_{l m}(k).
\label{RecWavelet}
\end{equation}

\bigskip

Equations (\ref{RecWavelet}) et (\ref{RecSmooth}) which define the wavelet decomposition are in fact equivalent to convolving the resolution at a given scale $j$ with a low-pass and a high-pass filter in order to obtain respectively the resolution and the wavelet coefficients at scale $j+1$.

The low-pass filter, denoted here by $h^j$, can be defined for each scale $j$ by :
\begin{equation}
\hat{h}^j_{l m} (k)  =  \left\lbrace \begin{array}{ll}
					 \frac{\hat{\Phi}^{2^{-(j+1)} k_c}_{0 0}(k)}{\hat{\Phi}^{2^{-j}k_c}_{0 0}(k) } & \mbox{ if $k < \frac{k_c}{2^{j+1}}$ and $l=m=0$,} \\
					0 & \mbox{ otherwise.}
					\end{array} \right.
\end{equation}

Then the approximation at scale $j+1$ is given by the convolution of scale $j$ with $h^j$ :
\begin{equation}
c^{j+1} = c^{j} \ast \frac{1}{\sqrt{2} \pi} h^j.
\end{equation}

In the same way, a high pass filter $g^j$ can be defined on each scale $j$ by:
\begin{eqnarray}
\hat{g}^j_{l m} (k) = \left\lbrace \begin{array}{ll}
					 \frac{\hat{\Psi}^{2^{-(j+1)} k_c}_{0 0}(k)}{\hat{\Phi}^{2^{-j}k_c}_{0 0}(k) } & \mbox{ if $k < \frac{k_c}{2^{j+1}}$ and $l=m=0$,} \\
					1 & \mbox{ if $k \geq \frac{k_c}{2^{j+1}}$ and $l=m=0$,} \\
					0 & \mbox{ otherwise.}
					\end{array} \right.
\end{eqnarray}
Given the definition of $\Psi$, $g^j$ can also be expressed in the simple form :
\begin{equation}
\hat{g}^j_{l m} (k) = 1 - \hat{h}^j_{l m} (k).
\end{equation}
The wavelet coefficients at scale $j$ are obtained by convolving the resolution at scale $j-1$ with $g^{j-1}$:
\begin{equation}
w^j = c^{j-1} \ast \frac{1}{\sqrt{2} \pi} g^{j - 1}.
\end{equation}

\bigskip

In summary, the two relations necessary to recursively define the wavelet transform are:
\begin{eqnarray}
\hat{c}^{j+1}_{l m}(k) & = & \hat{h}^j_{0 0}(k) \hat{c}^{j}_{l m}(k) ,\\
\hat{w}^{j+1}_{l m}(k) & = & \hat{g}^j_{0 0}(k) \hat{c}^{j}_{l m}(k).
\end{eqnarray}

\subsection{Reconstruction}

Since the wavelet coefficients are defined as the difference between two resolutions, the reconstruction from the wavelet decomposition $\mathcal{W} = \{w^1, \ldots, w^J, c^J\}$ is straightforward and corresponds to the reconstruction formula of the \textit{\`a trous} algorithm:
\begin{equation}
c^0(r,\theta_r,\phi_r) = c^J(r,\theta_r,\phi_r) + \sum_{j=1}^{J} w^j(r,\theta_r,\phi_r).\end{equation}
However, given the redundancy of the transform, the reconstruction is not unique. It is possible to take advantage of this redundancy to reconstruct $c^j$ from $c^{j+1}$ and $w^{j+1}$ by using a least squares estimate.

\bigskip

From the previous paragraph, the wavelet decomposition can be then recursively defined by:
\begin{eqnarray}
\hat{c}^{j+1}_{l m}(k) & = & \hat{h}^j_{0 0}(k) \hat{c}^{j}_{l m}(k) ,\\
\hat{w}^{j+1}_{l m}(k) & = & \hat{g}^j_{0 0}(k) \hat{c}^{j}_{l m}(k).
\end{eqnarray}
If these relations are respectively multiplied by $\overline{\hat{h}^j_{l m}}(k)$ and $\overline{\hat{g}^j_{l m}}(k)$ and then added together, the following expression is obtained  for the least squares estimate of $c^j$ from $c^{j+1}$ and $w^{j+1}$: 
\begin{equation}
\hat{c}^j_{l m} (k) = \hat{c}^{j+1}_{l m}(k) \hat{\tilde{h}}^j_{l m}(k)  + \hat{w}^{j+1}_{l m} \hat{\tilde{g}}^j_{l m}(k),
\end{equation}
where $\hat{\tilde{h}}^j$ and $\hat{\tilde{g}}^j$ are defined as follows:
\begin{eqnarray}
\hat{\tilde{h}}^j_{l m}(k) & = & \frac{\overline{\hat{h}^j}_{l m}(k)}{ |\hat{h}^j_{l m}(k)|^2 + |\hat{g}^j_{l m}(k)|^2 } ,\\
\hat{\tilde{g}}^j_{l m}(k) & = & \frac{\overline{\hat{g}^j}_{l m}(k)}{ |\hat{h}^j_{l m}(k)|^2 + |\hat{g}^j_{l m}(k)|^2 }.
\end{eqnarray}

\bigskip

Among the advantages of using this reconstruction formula instead of the raw sum over the wavelet coefficients is that there is no need to perform an inverse and then direct SFB transform to reconstruct the coefficients of the original data. Indeed, both the wavelet decomposition and reconstruction procedures only require access to SFB coefficients and there is no need to revert back to the direct space (although this will be required if one wishes to observe the power of each coefficient in real space as we do in Section \ref{section:WaveletTest} for denoising purposes).

\subsection{Choice of a scaling function}

Any function with spherical symmetry and a cut-off frequency $k_c$ would do as a scaling function but in this work we choose to use a B-spline function of order 3 to define our scaling function:

\begin{equation}
\hat{\Phi}^{k_c}_{l m}(k) = \frac{3}{2} B_3\left( \frac{2 k }{k_c} \right) \delta_{l 0} \delta_{m 0},
\end{equation}
where
\begin{equation}
B_3(x) = \frac{1}{12}\left( | x -2 |^3 - 4 | x - 1 |^3 + 6 |x|^3 - 4 |x +1|^3 + |x +2|^3 \right).
\end{equation}

\begin{figure*}[htb]
\begin{center}
\begin{tabular}{c|c}
\subfigure[Scaling function $\hat{\Phi}^{2^{-j} k_c}_{0 0}(k)$ for $j=0,1,2$]{\includegraphics[width=0.45\textwidth]{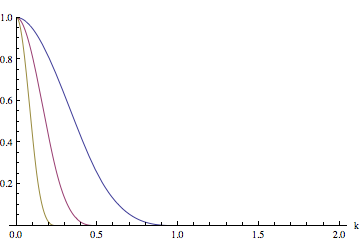}}&
\subfigure[Wavelet function $\hat{\Psi}^{2^{-j} k_c}_{0 0}(k)$ for $j=0,1,2$]{\includegraphics[width=0.45\textwidth]{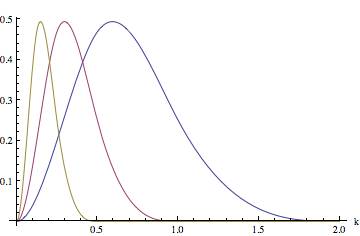}} 
\end{tabular}
\end{center}
\caption{Scaling function and Wavelet function for $k_c=1$. The $y$-axis represents the amplitude and the $x$-axis the frequency associated with the scaling function.  The units of the amplitude must be those of the SFB transform.}
\label{Scaling_and_Wavelet_Functions}
\end{figure*}

The scaling function and its corresponding wavelet function are plotted in SFB space for different values of $j$ in Figure \ref{Scaling_and_Wavelet_Functions}. 

\begin{figure*}[htb]
\centerline{
\hbox{
 \includegraphics[width=7.5cm, trim= 2cm 14cm 2cm 2cm, clip=]{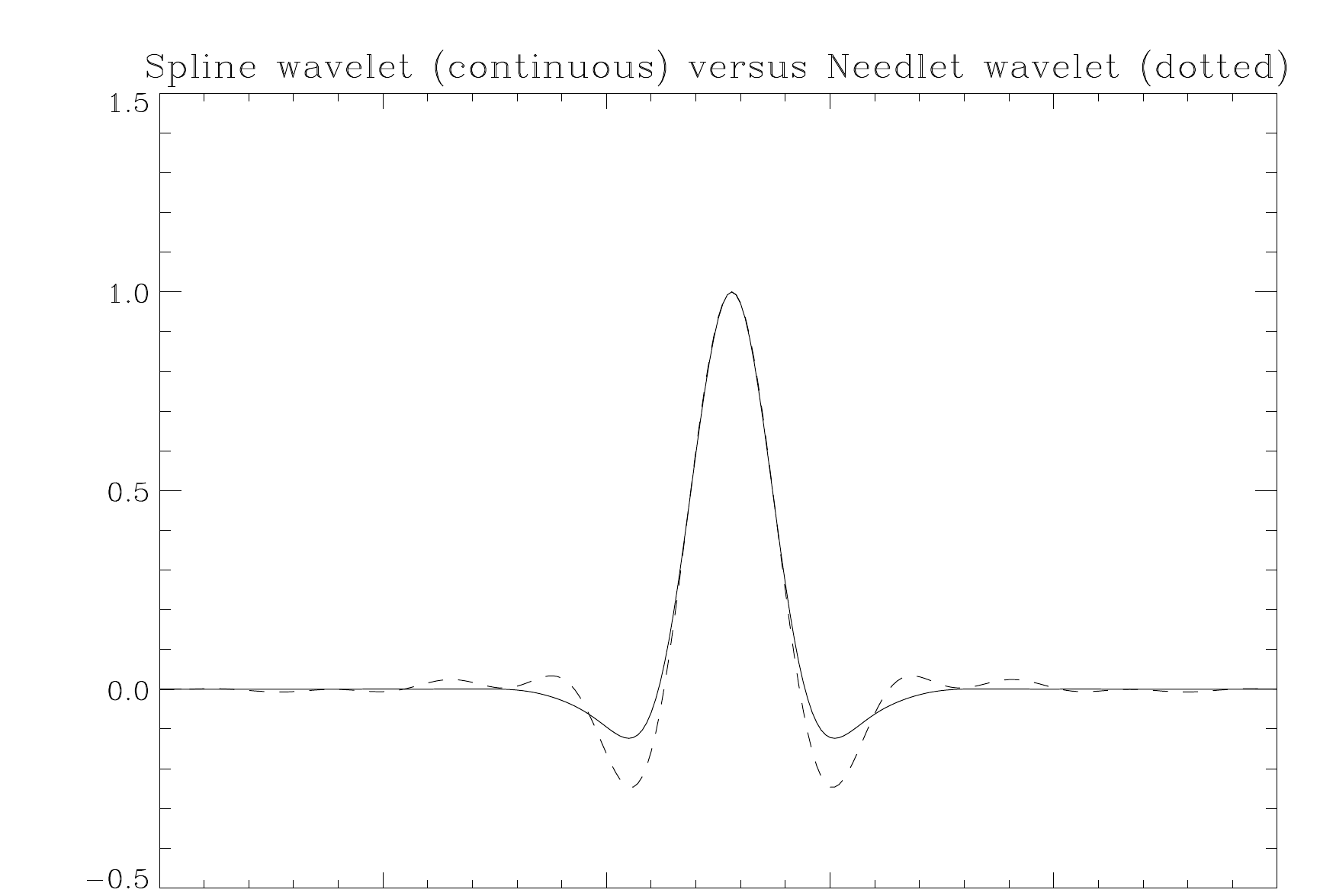}
 \includegraphics[width=7.5cm, trim= 2cm 14cm 2cm 2cm, clip=]{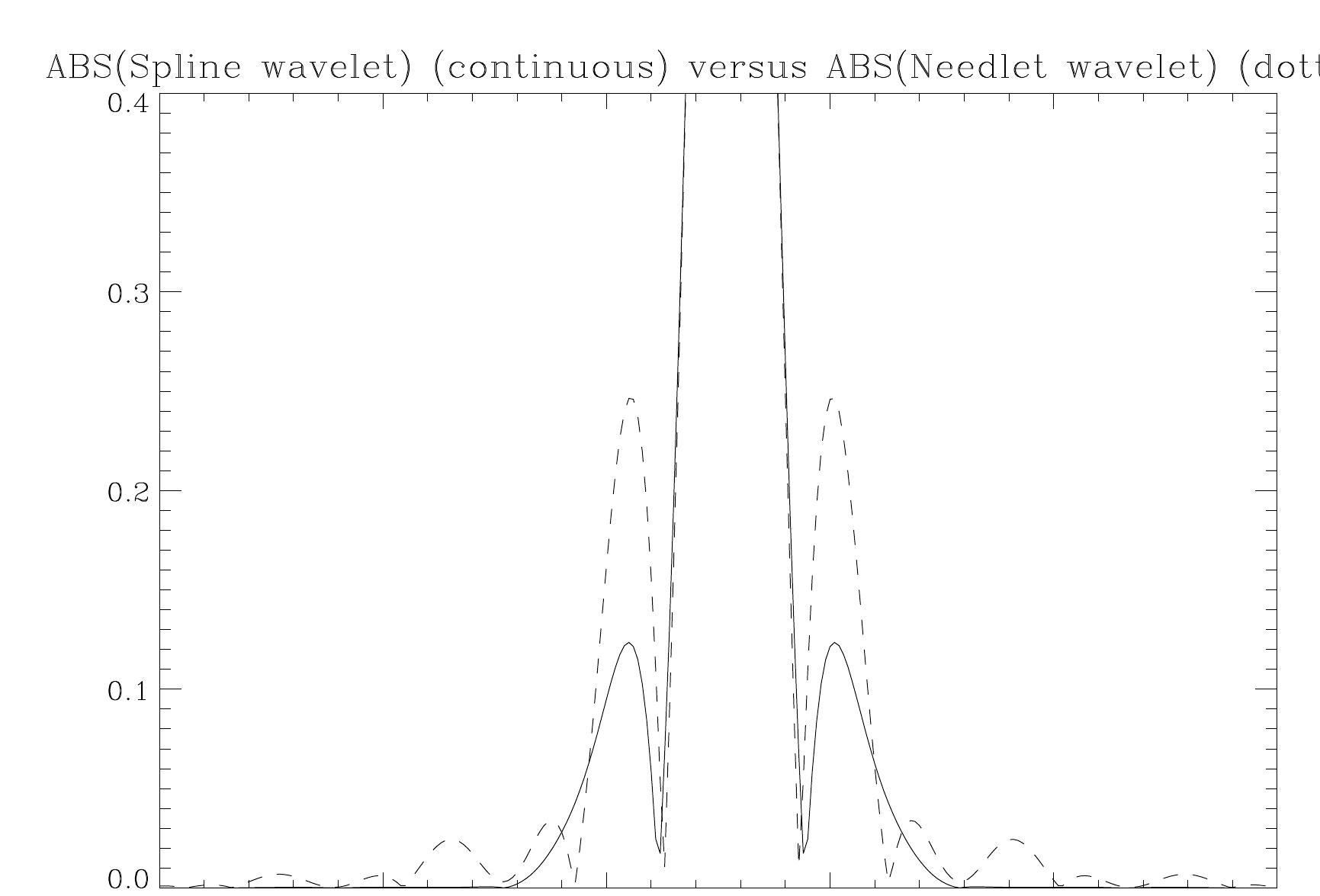}
}}
\caption{Comparaison between spline and needlet wavelet functions on the sphere. The $y$-axis corresponds to amplitude and the $x$-axis to position.}
\label{Figure:cmp_needlet}
\end{figure*}

Other functions such as Meyer wavelets or the needlet function \citep{marinucci08} can be used as well. 
Needlet wavelet functions have a much better frequency localisation than the wavelet function derived from the  B$_3$-spline but present more oscillations in the direct space. To illustrate this, we 
show in Fig.~\ref{Figure:cmp_needlet} two different wavelet functions. Fig.~\ref{Figure:cmp_needlet}~left   shows  1D profile of the spline  (continuous line) 
and the needlet wavelet function (dotted line) at a given scale. Fig.~\ref{Figure:cmp_needlet}~right   shows the same function,  
but we have plotted the absolute value in order to better visualise their respective ringing.
As it can be seen, for wavelet functions with the same main lob, the needlet wavelet oscillates much more than the spline wavelet. 
Hence, the best wavelet choice certainly depends on the desired applications. For statistical analysis, detection or restoration applications, 
we may prefer to use a wavelet which does not oscillate too much and with a smaller support, in this case the spline wavelet is clearly the 
correct choice. For spectral or bispectral analysis, where the frequency localisation is fundamental, then needlets should be 
preferred to the spline wavelet.

\section{The Discrete Spherical Fourier-Bessel Transform (DSFBT)}\label{sec:DSFBT}

The spherical 3D wavelet formalism derived in the previous section is based on a continuous SFB transform. However, a continuous transform is not practical numerically and a discrete equivalent is required. In the case of galaxy surveys, where the underlying density field is sampled at discrete points, this problem can be addressed by combining a boundary condition with the discrete nature of the survey \citep[see for e.g. Equations 5 and 7 in][even though they use a slightly different SFB expansion as in this paper]{Leistedt2011} which yields in this specific case a practical way to estimate discrete SFB coefficients. However this method cannot be used for the purpose of performing wavelet treatments since this requires a continuous field. 

Indeed, to perform operations on the wavelet scales (e.g. for denoising purposes), one first needs to recover the wavelet coefficients in physical space, apply a treatment to those coefficients (e.g. thresholding) and then convert them back into SFB space (as required by the  wavelet transform algorithm which operates in SFB space). This last operation requires a SFB transform which takes as input fields and not discrete galaxy surveys.
 
 In this section, we show how a discretisation of the k spectrum in combination with a Healpix type angular pixelisation, can be used to build a fast Discrete SFB Transform (DSFBT) which allows for a novel and practical two-way conversion between the discrete SFB coefficients and discrete values of the field in physical space sampled on a spherical 3D grid.

\subsection{The Discrete Spherical Bessel Transform}\label{sub:DSBT}

The discretisation of the SFB transform along the radial component uses the well known orthogonality property of the Spherical Bessel functions on the interval $[0, R]$. If $f$ is a continuous function defined on $[0, R]$  which verifies the boundary condition $f(R) = 0$ then the SBT defined by Eq. (\ref{Def_SBT}) can be expressed using SFB series:
\begin{eqnarray}
	\hat{f}_l(k_{l n}) & = & \sqrt{\frac{2}{\pi}} \int_0^R f(r) j_{l} ( k_{l n} r) r^2 dr \label{Discrete_transform_r},\\
	f(r)    & = & \sum\limits_{n = 1}^{\infty} \hat{f}_l(k_{l n})  \rho_{l n} j_{l} (k_{l n} r)  \label{Discrete_inverse_r}.
\end{eqnarray}
In this expression,  $k_{l n} = \frac{q_{l n}}{R}$ where $q_{l n}$ is the nth zero of the Bessel function of order $l$ and the weights $\rho_{l n}$ are defined as:
\begin{equation}
	\rho_{l n} = \frac{\sqrt{2 \pi} R^{-3}}{ j_{l+1}^2(q_{l n})}.
\end{equation}

	Although this formulation provides a discretisation of the inverse SBT and of the $k$ spectrum, the direct transform is still continuous and another discretisation step is required.
	
	If a boundary condition of the same kind is applied to $\hat{f}_l(k)$ so that $\hat{f}_l(K_l) =0$, then by using the same result, the SFB expansion of $\hat{f}_l(k)$ is obtained by:
	\begin{subequations}
	\begin{align}
	\hat{\hat{f}}_l(r_{l n}) & = & \sqrt{\frac{2}{\pi}} \int_0^K \hat{f}_l(k) j_{l} ( r_{l n} k) k^2 dk \label{Discrete_transform_k},\\
	\hat{f}_l(k)    & = & \sum\limits_{n = 1}^{\infty} \hat{\hat{f}}_l(r_{l n})  \kappa_{l n} j_{l} (r_{l n} k)  \label{Discrete_inverse_k},
\end{align}
\end{subequations}
	where $r_{l n} = \frac{q_{l n}}{K_l}$ and where the weights $\kappa_{l n}$ are defined as: 	
	\begin{equation}
	\kappa_{l n} = \frac{\sqrt{2 \pi} K_l^{-3}}{ j_{l+1}^2(q_{l n})}.
\end{equation}
	
	The SBT being an involution, $\hat{\hat{f}} = f$ so that $\hat{\hat{f}}_l(r_{l n}) = f(r_{l n})$. Much like the previous set of equations had introduced a discrete $k_{l n}$ grid, a discrete $r_{l n}$ grid is obtained for the radial component. Since Eqs. (\ref{Discrete_inverse_r}) and (\ref{Discrete_inverse_k}) can be used to compute $f$ and $\hat{f}_l$ for any value of $r$ and $k$, they can in particular be used to compute $f(r_{l n})$ and $\hat{f}_l(r_{l' n})$ where $l'$ does not have to match $l$. The SBT and its inverse can then be expressed only in terms of series:
	
\begin{subequations}
	\begin{align}
 \hat{f}_l(k_{l' n})  &=& \sum\limits_{p = 1}^{\infty} f(r_{l p})  \kappa_{l p} j_{l} (r_{l p} k_{l' n}), \label{Direct_series}\\
 f(r_{l' n}) &=& \sum\limits_{p = 1}^{\infty} \hat{f}_l(k_{l p})  \rho_{l p} j_{l} (r_{l' n} k_{l p}).  \label{Inverse_series}
\end{align}
\end{subequations}
	With these equations one can easily compute the SBT and its inverse without the need of evaluating any integral. Furthermore only discrete values of $f$ and $\hat{f}$ respectively sampled on $r_{l n}$ and $k_{l n}$ are required.

	In practical applications, for a given value of $l$ only a limited number $N_{max}$ of $\hat{f}_l(k_{l n})$ and $f(r_{l n})$ coefficients can be stored so that $r_{l N_{max}} = R$ and $k_{l N_{max}} = K_l$. Since $r_{l n}$ is defined by $r_{l n} = \frac{q_{l n}}{K_l}$, for $n = N_{max}$ $R$ and $K_l$ are bound by the following relationship:
	\begin{equation}
 q_{l N}		   =  K_l R.
 \end{equation}
 Therefore, from the value of $R$ one can easily determine the appropriate value for $K_l$. Furthermore, because of the boundary conditions on $\hat{f}_l$ and  $f$, the series (\ref{Direct_series}) and (\ref{Inverse_series}) can now be truncated to $N_{max}$ terms. 
 
 This truncation allows the use of a very convenient matrix formalism to represent the transform. In their explicit form, the two truncated sums can be rewritten as:

\begin{eqnarray}
 \hat{f}_l(k_{l' n}) & = & K_l^{-3} \sum\limits_{p = 1}^{N_{max}} f(r_{l p})  \frac{\sqrt{2 \pi}}{ j_{l+1}^2(q_{l p})} j_{l} \left(\frac{q_{l p} q_{l' n}}{q_{l N_{max}}} \right) ,\\
 f(r_{l' n})  & = &  R^{-3} \sum\limits_{p = 1}^{N_{max}} \hat{f}_l(k_{l p})  \frac{\sqrt{2 \pi}}{ j_{l+1}^2(q_{l p})} j_{l} \left(\frac{q_{l p} q_{l' n}}{q_{l N_{max}}}\right) .
\end{eqnarray}
From these equations, a transform matrix $T^{l l'}$ can be defined as:
\begin{equation}
	T_{p q}^{l l'} = \left( \frac{\sqrt{2 \pi}}{ j_{l+1}^2(q_{l q})} j_{l} (\frac{q_{l' p} q_{l q}}{q_{l N_{max}}}) \right)_{p q}.
\end{equation}
This matrix allows the computation of $\hat{f}_l$ on any grid $k_{l' n}$ from the values of $f$ sampled on $r_{l n}$:
\begin{equation}
	\left[\begin{matrix}
	\hat{f}_l(k_{l' 1}) \\
	\hat{f}_l(k_{l' 2}) \\
	\vdots \\
	\hat{f}_l(k_{l' N_{max}})
	\end{matrix}\right] = \frac{1}{K_l^3} T^{l l'} \left[\begin{matrix}
	f(r_{l 1}) \\
	f(r_{l 2}) \\
	\vdots \\
	f(r_{l N_{max}})
	\end{matrix}\right] .\label{Direct_DSBT}
\end{equation}
Reciprocally, the inverse the values of $f$ can be computed on any $r_{l' n}$ grid from $\hat{f}_l$ sampled on $k_{l' n}$ using the exact same matrix:
\begin{equation}\left[\begin{matrix}
	f(r_{l' 1}) \\
	f(r_{l' 2}) \\
	\vdots \\
	f(r_{l' N})
	\end{matrix}\right] = \frac{1}{R^3} T^{l l'} \left[\begin{matrix}
	\hat{f}_l(k_{l 1}) \\
	\hat{f}_l(k_{l 2}) \\
	\vdots \\
	\hat{f}_l(k_{l N})
	\end{matrix}\right] .\label{Inverse_DSBT}
\end{equation}

The discrete SBT is defined by the set of equations (\ref{Direct_DSBT}) and (\ref{Inverse_DSBT}). 

A simplified form of the transform could have been defined only for $l = l'$ so that $T^{l l'} = T^{l}$. However, keeping the distinction between the order of the transform $l$ and the order of the grid on which the results are provided $l'$ will be crucial to the implementation of the DSFBT. Indeed, the order of the grid on which the function is sampled has to match the order of the transform but the resulting transform coefficients do not. Therefore, it will be possible to compute the result of the inverse SBT of any order $l$ on a grid of order $l_0$ so that only one radial grid of order $l_0$ will be required. Nevertheless, for the direct transform, if the field is sampled on the radial grid of order $l_0$, only the transform of order $l_0$ can be computed. An additional result is required to be able to relate the SBT of different orders between them. This is achieved by combining relations (\ref{Discrete_transform_r}) and (\ref{Discrete_inverse_r}):

\begin{eqnarray}
	\hat{f}_l(k_{l n})  & = & \sqrt{\frac{2}{\pi}} \int_0^R \left[ \sum\limits_{m = 1}^{\infty} \hat{f}_{l_0}(k_{l_0 m})  \rho_{l_0 m} j_{l_0} (k_{l_0 m} r) \right] j_{l} ( k_{l n} r) r^2 dr \nonumber ,\\
				   & = & \sqrt{\frac{2}{\pi}} \sum\limits_{m = 1}^{\infty} \hat{f}_{l_0}(k_{l_0 m}) \rho_{l_0 m} \int_0^R j_{l_0} (k_{l_0 m} r) j_{l} ( k_{l n} r) r^2 dr \nonumber ,\\
				   & = & \sum\limits_{m = 1}^{\infty} \hat{f}_{l_0}(k_{l_0 m}) \frac{2}{ j_{l_0+1}^2(q_{l_0 m})} \int_0^1 j_{l_0} (q_{l_0 m} x) j_{l} ( q_{l n} x) x^2 dx \nonumber ,\\
				   & = & \sum\limits_{m = 1}^{\infty} \hat{f}_{l_0}(k_{l_0 m}) \frac{2}{ j_{l_0+1}^2(q_{l_0 m})} W_{n m}^{l_0 l} ,\label{TransformConversion}
\end{eqnarray}
where the weights $W_{n m}^{l l'}$ are defined as:
\begin{equation}
	W_{n m}^{l_0 l} = \int_0^1 j_{l_0} (q_{l_0 m} x) j_{l} ( q_{l n} x) x^2 dx.
	\label{Transform_weigths}
\end{equation}

	The final expression in Equation \ref{TransformConversion} is an important result which shows that the SBT of a given order can be expressed as the sum of the coefficients obtained for a different order of the transform, with the appropriate weighting. This means we can convert the Spherical Bessel coefficients of order $l_0$ into coefficients of any other order $l$, which considerably speeds up calculations for the SBT. It is also worth noticing that the weights $W_{n m}^{l l'}$ are simply geometric terms, i.e. independent of the field and can thus be tabulated.

	We note that this approach is an extension of the Discrete SBT presented in \cite{Lemoine:1994} using spherical Bessel functions and where the transform in \cite{Lemoine:1994} can be considered as a special case where $l=l'$. 

\subsection{A Spherical 3D grid compatible with a Discrete Spherical Fourier-Bessel spectrum}

As presented in section \ref{section:SFBT}, the SFB transform is the composition of a SHT for the angular component and a SBT for the radial component. Since these two transforms can commute, they can be treated independently and by combining discrete algorithms for both transforms, one can build a DSFBT. In this work, the angular part of the transform is implemented using the HEALPix pixelisation scheme while the radial component uses the discrete SBT algorithm presented in Section \ref{sub:DSBT}. The choice of these two algorithms introduces a discretisation of the SFB coefficients as well as a 3D gridding of the density in physical space. 

\bigskip

The SFB coefficients $\hat{f}_{l m}(k)$ are defined by Equation ($\ref{SFBT}$) for continuous values of $k$. Assuming a boundary condition on the density field $f$, the discrete SBT can be used to discretise the values of $k$. The Discrete SFB coefficients are therefore defined as:
\begin{equation}
f_{l m n} \equiv \hat{f}_{l m}(k_{l n}),
\end{equation}
for $0 \leq l \leq L_{max}$, $-l \leq m \leq l$ and $1 \leq n \leq N_{max}$. These discrete coefficients are simply obtained by sampling the continuous coefficients on the $k_{l n}$ grid introduced in the previous section. 

\bigskip

To this discretised SFB spectrum corresponds a dual grid of the 3D space defined by combining the HEALPix pixelisation scheme and the discrete SBT. 

Used for the computation of the SHT, the HEALPix scheme maps the sphere using curvilinear quadrilaterals of different shapes but equal area. Although other pixelisation schemes for the sphere such as IGLOO or GLESP could have been used, the choice of HEALPix is essentially motivated by the homogeneity of the pixelisation and by its comprehensive software package. For a given value of $r$, the field $f(r,\theta, \phi)$ can therefore be sampled on a finite number of points using HEALPix.

The radial component of the transform is conveniently performed using the discrete SBT. Indeed, this algorithm introduces a radial grid compatible with the discretised $k_{l n}$ spectrum. Although this radial grid $r_{l n}$ depends on the order $l$ of the SBT, it will be justified in the next section that only one grid $r_{l_0 n}$ is required to sample the field along the radial dimension. The value of $l_0$ is set to 0 because in this case the properties of the zeros of the Bessel function ensure that $r_{0 n}$ will be regularly spaced between 0 and $R$:
\begin{equation}
r_{0 n} = \frac{n}{N_{max}} R.
\end{equation}
For given values of $\theta$ and $\phi$, the field $f(r,\theta,\phi)$ can now be sampled on discrete values of $r = r_{0 n}$.

\bigskip

\begin{figure}[hbt]
\begin{center}
\includegraphics[width=0.45\textwidth]{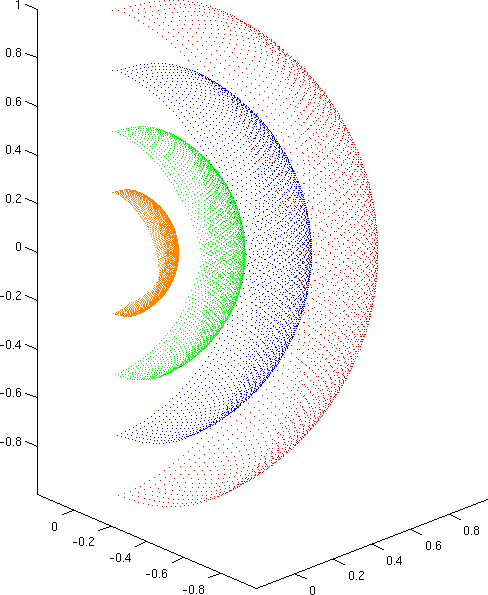}
\caption{Representation of the spherical 3D grid for the DSFBT ($R=1$ and $N_{max} = 4$) as described in Section \ref{sec:DSFBT} where the axes represent $x, y, z$ and the grid points correspond to the centre of Healpix pixels and the radial grid is defined in Section \ref{sub:DSBT}.}
\label{Spherical_3D_grid} 
\end{center}
\end{figure}

Combining angular and radial grids, the 3D spherical grid is defined as a set of $N_{max}$ HEALPix maps equally spaced between 0 and $R$. An illustration of this grid is provided on Fig.\ref{Spherical_3D_grid} where only one quarter of the space is represented for clarity. 
		
With this 3D grid, it will be possible to compute back and  forth the SFB transform between a density field and its SFB coefficients. The details of the actual algorithm are provided in the next section.

\subsection{Algorithm for the Discrete Spherical Fourier-Bessel Transform (DSFBT)}
	
	Although the SHT and the SBT formally commute, some practical considerations on the order between the two operations are to be taken into account for their actual implementation. Here, a detailed algorithm of both the direct and inverse DSFBT is provided.

	\subsubsection{Direct Transform} Given a density field $f$ sampled on the spherical 3D grid, the SFB coefficients $f_{l m n}$ are computed in three steps:
	
	\begin{itemize}
		\item[1)] For each $n$ between 1 and $N_{max}$ the SHT of the HEALPix map of radius $r_{l_0 n}$ is computed. This yields $f_{l m} (r_{l_0 n})$ coefficients.

		\item[2)] The next step is to compute the SBT of order $l_0$ from the $f_{l m} (r_{l_0 n})$ coefficients for every $(l, m)$. This operation is a simple matrix product:
\begin{equation}
	\left\lbrace\begin{matrix}\forall & 0 &\leq & l & \leq & L_{max} \\ \forall & -l & \leq & m & \leq & l\end{matrix} \right., 
	\left[\begin{matrix}
	\hat{f}^{l_0}_{l m}(k_{l_0 1}) \\
	\hat{f}^{l_0}_{l m}(k_{l_0 2}) \\
	\vdots \\
	\hat{f}^{l_0}_{l m}(k_{l_0 N_{max}})
	\end{matrix}\right]
	 = \frac{T^{l_0 l_0}}{K^3}  \left[\begin{matrix}
	f_{l m}(r_{l_0 1}) \\
	f_{l m}(r_{l_0 2}) \\
	\vdots \\
	f_{l m}(r_{l_0 N_{max}})
	\end{matrix}\right].
\end{equation}
	This operation yields $\hat{f}^{l_0}_{l m}(k_{l_0 n})$ coefficients which are not yet SFB coefficients because the order of the SBT $l_0$ does not match the order of the Spherical Harmonics coefficients $l$. An additional step is required.
	
	\bigskip
	
	\item[3) ] The last step required to gain access to the SFB coefficients $f_{l m n}$ is to convert the Spherical Bessel coefficients for order $l_0$ to the correct order $l$ that matches the Spherical Harmonics order. This is done by using relation (\ref{TransformConversion}):
	
	\begin{eqnarray}
	\left\lbrace\begin{matrix}\forall & 0 &\leq & l & \leq & L_{max} \\ \forall & -l & \leq & m & \leq & l \\ \forall & 1 & \leq & n & \leq & N_{max}\end{matrix} \right., \hat{f}_{l m}(k_{l n}) = \sum\limits_{p = 1}^{N_{max}} \hat{f}^{l_0}_{l m}(k_{l_0 p}) \frac{2 W_{n p}^{l_0 l} }{ j_{l_0+1}^2(q_{l p})}
,\end{eqnarray}
where $W_{n p}^{l_0 l}$ are defined by Eq. (\ref{Transform_weigths}).
This operation finally yields the $f_{l m n} = \hat{f}_{l m}(k_{l n})$ coefficients.	

\end{itemize}
	
		\subsubsection{Inverse Transform} Let $f_{l m n}$ be the coefficients of the SFB transform of the density field $f$, where the $f_{l m n}$ coefficients can be calculated for a continuous density field using the direct transform described above or for a galaxy or halo catalogue using the public code 3DEX \citep{Leistedt2011}. Note that the 3DEX code can account for masked regions of missing data. The reconstruction of $f$ on the spherical 3D grid requires two steps:
		
		\begin{itemize}
		\item[1)]	First, from the $f_{l m n}$, the inverse discrete SBT is computed for all $l$ and $m$. Again, this transform can be easily evaluated using a matrix product:
	\begin{equation}
	\left\lbrace\begin{matrix}\forall & 0 &\leq & l & \leq & L_{max} \\ \forall & -l & \leq & m & \leq & l\end{matrix} \right., \quad
	\left[\begin{matrix}
	f_{l m}(r_{l_0 1}) \\
	f_{l m}(r_{l_0 2}) \\
	\vdots \\
	f_{l m}(r_{l_0 N_{max}})
	\end{matrix}\right] = \frac{T^{l l_0}}{R^3}  \left[\begin{matrix}
	f_{l m 1} \\
	f_{l m 2} \\
	\vdots \\
	f_{l m N_{max}}
	\end{matrix}\right].
\end{equation}
	Here, it is worth noticing that the matrix $T^{l l_0}$ allows the evaluation of the SBT of order $l$ and provides the results on the grid of order $l_0$. 
	\bigskip	
	\item[2)] From the spherical harmonics coefficients $f_{l m}(r_{l_0 n})$ given at specific radial distances $r_{l_0 n}$ it is possible to compute the inverse SHT. For each $n$ between 1 and $N_{max}$ the HEALPix inverse SHT is performed on the set of coefficients $\{ f_{l m}(r_{l_0 n}) \}_{l,m}$. This yields $N_{max}$ HEALPix maps which constitute the sampling of the reconstructed density field on the 3D spherical grid.
	
	\end{itemize}

\section{Wavelet decomposition of a test density field}
\label{section:WaveletTest}

	To illustrate the wavelet transform described in Section \ref{sec:wavelet}, a set of SFB Coefficients was extracted from a 3D density field using the DSFBT described in in Section \ref{sec:DSFBT}. The test density field was provided by a cosmological N-body simulation which was carried out by the Virgo Supercomputing Consortium using computers based at Computing Centre of the Max-Planck Society in Garching and at the Edinburgh Parallel Computing Centre. We use their large box simulations\footnote{\url{http://www.mpa-garching.mpg.de/Virgo/VLS.html}, a $\Lambda {\rm CDM}$ simulation at $z=0$, which was calculated using $512^3$ particles for the following cosmology : $\Omega_m=0.3, \Omega_\Lambda=0.7, H_0=70km{\rm s}^{-1}{\rm Mpc}^{-1}, \sigma_8=0.9$. The data cube provided is $479 h^{-1}{\rm Mpc}$ in length}.  
	
\begin{figure}[hbtp]
\begin{center}
\begin{tabular}{cc}
\subfigure[Reconstructed density from the initial SFB coefficients]{\includegraphics[width=0.22\textwidth]{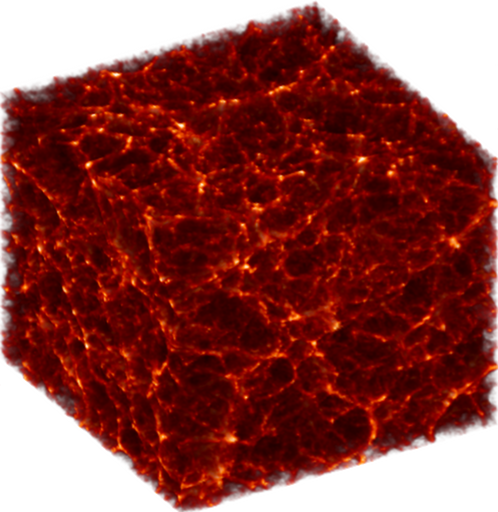}\label{Wavelet_decompositiona}} & \subfigure[First wavelet scale]{\includegraphics[width=0.22\textwidth]{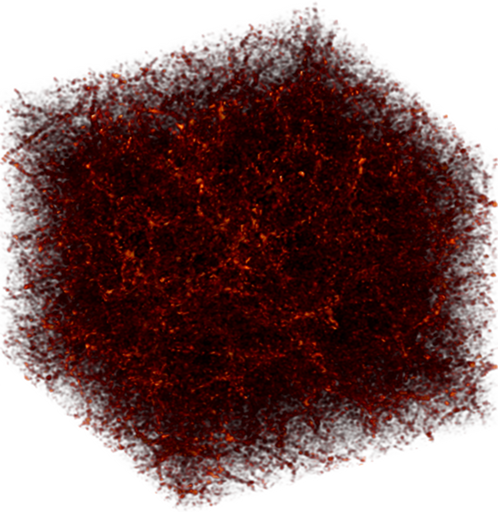}\label{Wavelet_decompositionb}} \\
\subfigure[Second wavelet scale]{\includegraphics[width=0.22\textwidth]{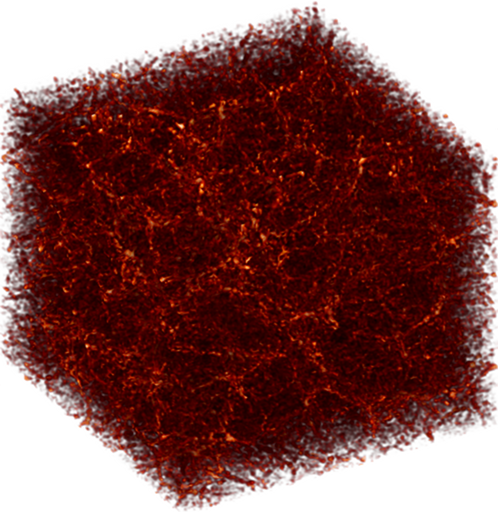}\label{Wavelet_decompositionc}} & \subfigure[Third wavelet scale]{\includegraphics[width=0.22\textwidth]{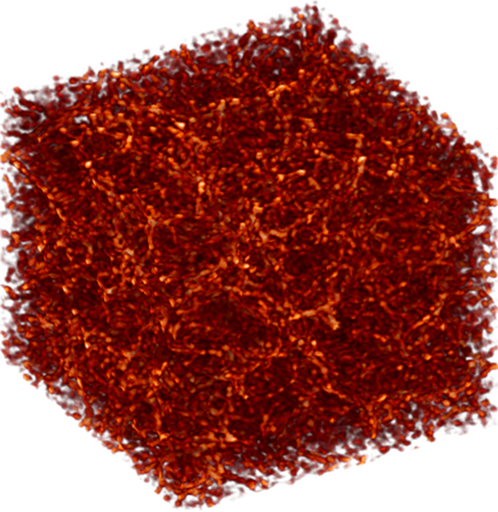}\label{Wavelet_decompositiond}} \\
\subfigure[Fourth wavelet scale]{\includegraphics[width=0.22\textwidth]{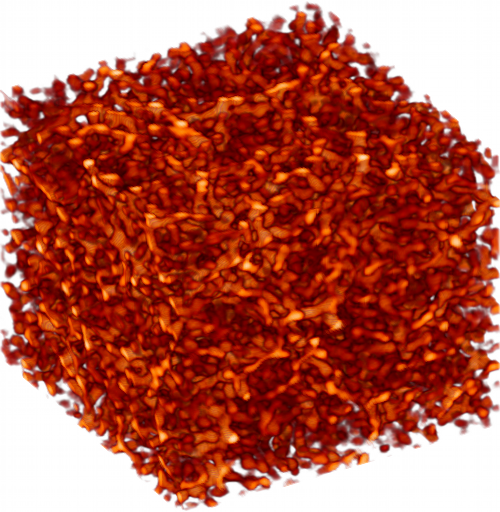}\label{Wavelet_decompositione}} & \subfigure[Smoothed density]{\includegraphics[width=0.22\textwidth]{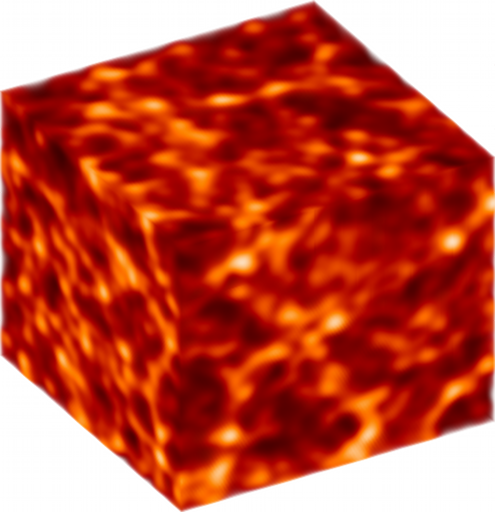}\label{Wavelet_decompositionf}}
\end{tabular}
\caption{Spherical 3D wavelet decomposition of a density field for a box with side $239.5 h^{-1}{\rm Mpc}$. Panel (a) shows the reconstructed density field directly using the initial SFB coefficients. Panels (b)-(e) show the first four wavelet scales, while panel (f) shows the smoothed density field.}
\label{Wavelet_decomposition}
\end{center}
\end{figure}

	We first compute the SFB coefficients of the test density field by sampling the Virgo density field on the 3D grid illustrated in Figure \ref{Spherical_3D_grid}, for $nside=2048$, $l_{\rm max}=1023$ and $n_{\rm max}=512$. In order to perform the SFB decomposition, we place ourselves in the middle of the box, and calculate the SFB coefficients out to $r = 479/2~ h^{-1}{\rm Mpc}$, setting the over-density field to zero outside of this spherical volume. In order to better visualise the data, we present in Figure \ref{Wavelet_decompositiona} the data in a cube spanning half the size of the original box (i.e. $239.5h^{-1}{\rm Mpc}$), where the field has been reconstructed from the SFB coefficients, using the discrete inverse transform we discuss in Section \ref{sec:DSFBT}.

From the SFB coefficients we calculate the wavelet decomposition, which yields the SFB coefficients of the various wavelet scales and the smoothed density field. Using the inverse DSFBT, the actual wavelet coefficients can be retrieved in the form of 3D density fields. These density fields corresponding to different wavelet scales are shown on Figure \ref{Wavelet_decompositionb}  to \ref{Wavelet_decompositione}. The smoothed density field which arises from the wavelet decomposition is also shown in Figure \ref{Wavelet_decompositionf}, which is simply given by: 
\begin{equation}(f) = (a) - (b) - (c) - (d) - (e) .\end{equation}
				
\section{Application to wavelet hard thresholding (denoising)}

In this section, a noise removal application based on wavelet thresholding is presented as an example of a potential use for the Isotropic Undecimated Spherical 3D Wavelet transform.

\begin{figure*}[hbt]
\begin{center}
\begin{tabular}{cc}
\subfigure[Initial density field]{\includegraphics[width=0.45\textwidth]{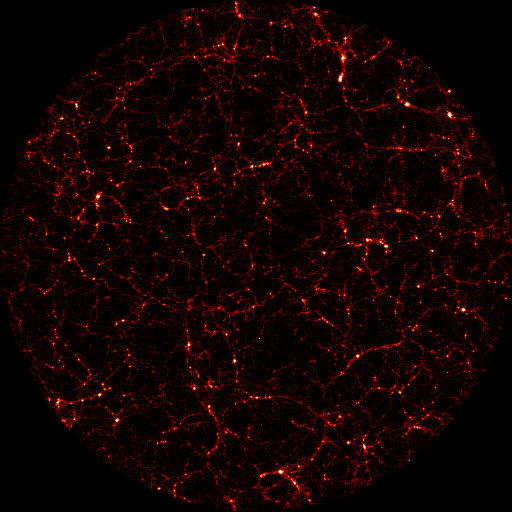} \label{Figure:fieldOriginal}} &
\subfigure[Noisy density field]{\includegraphics[width=0.45\textwidth]{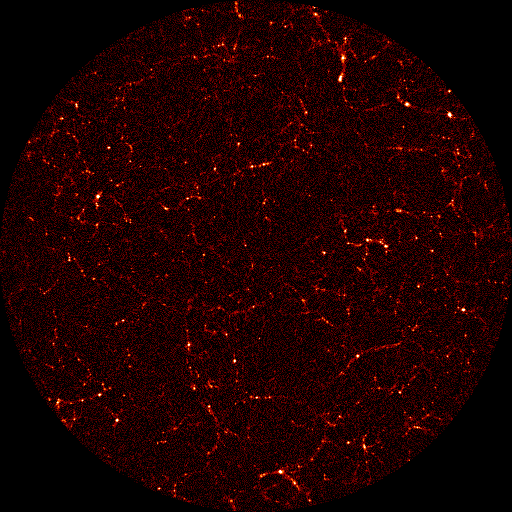}\label{Figure:fieldNoisy}} \\
\subfigure[Density field after wavelet hard thresholding]{\includegraphics[width=0.45\textwidth]{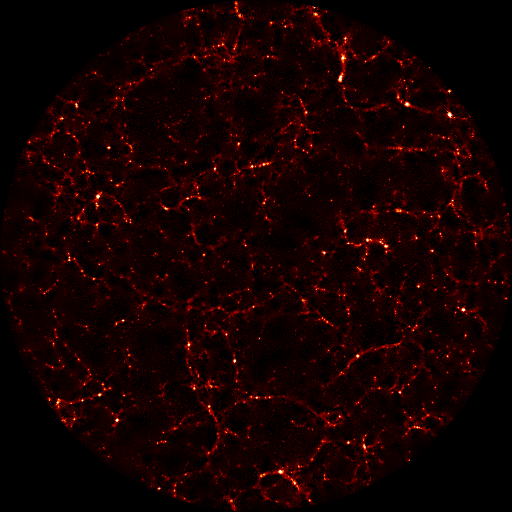} \label{Figure:fieldThresholded}} &
\subfigure[Residuals]{\includegraphics[width=0.45\textwidth]{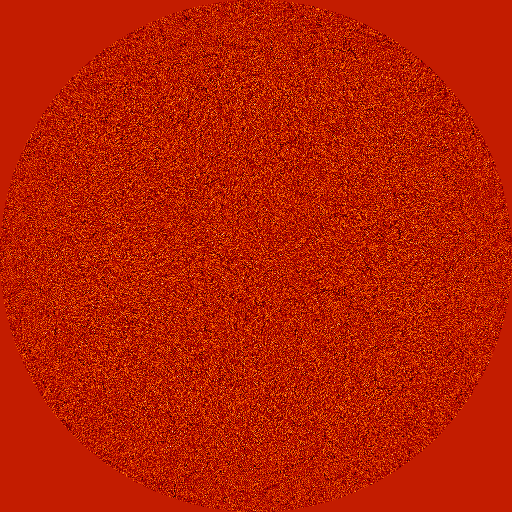}\label{Figure:residuals}} 
\end{tabular}
\caption{Wavelet Hard thresholding applied to a test density field}
\end{center}
\end{figure*}

\subsection{Thresholding}
Many wavelet filtering methods have been proposed in the last twenty years. {\em Hard thresholding} consists of setting to 0 all 
wavelet coefficients which have an absolute value lower than a threshold $T_j$ (non-significant wavelet coefficient):
\begin{eqnarray}  \tilde w_{j,k} = 
\left\{ \begin{array}{ll} w_{j,k} &  \mbox{ if } \mid w_{j,k} \mid \geq T_j  \nonumber  ,\\ 

0 &  \mbox{ otherwise,}  \end{array} \right. 
\end{eqnarray}
where $w_{j,k}$ is a wavelet coefficient at scale $j$ and at spatial position $k$. 

{\em Soft thresholding} consists of replacing each wavelet coefficient by the value $\tilde w$ where
\begin{eqnarray}  \tilde w_{j,k} = 
\left\{ \begin{array}{ll} sgn(w_{j,k}) ( \mid w_{j,k} \mid - T_j)    &  \mbox{ if } \mid w_{j,k} \mid \geq T_j \nonumber,  \\ 
0 &  \mbox{ otherwise}  .\end{array} \right. 
\end{eqnarray} 
This operation is generally written as:
\begin{eqnarray} 
 \tilde w_{j,k} = \mathrm{soft}( w_{j,k})  = sgn(w_{j,k}) ( \mid w_{j,k} \mid - T_j)_{+},
\end{eqnarray} 
where $(x)_{+} = \max(0,x)$.

In the case of Gaussian noise with a given standard deviation $\sigma$, a reasonable choice for the threshold  $T_j$ is  $ T_j = K \sigma_j$, where $j$ is the scale of the wavelet coefficient, 
$\sigma_j$ is the noise standard deviation at the scale $j$, and $K$ is a constant generally chosen between to 3 and 5.  The standard deviation $\sigma_j$ depends only on the chosen wavelet 
function and the noise level $\sigma$.  More details can be found in \citet{starck:book10}.

Other threshold methods have been proposed, like the {\em universal threshold} 
\citep{rest:donoho93_1,rest:donoho93_2}, or the SURE (Stein Unbiased Risk Estimate) method \citep{rest:donoho95},
but they generally do not yield as good results as the hard thresholding method based on the significant coefficients.  
Concerning the threshold level, the universal threshold  corresponds to a minimum risk. The larger the number of pixels, the larger the risk is, and it is normal that the threshold $T$ depends on the number of pixels ($T = \sigma_j \sqrt{2\log n}$, 
$n$ being the number of pixels). The $K\sigma$ threshold corresponds to a false detection probability, the probability 
to detect a coefficient as significant when it is due to the noise. The $3\sigma$ value corresponds to 0.27 \% false detection.
  
\subsection{Denoising experiment} We test the hard thresholding algorithm by using it to remove artificially added Gaussian noise on the Virgo density field (the simulation is described in Section \ref{section:WaveletTest}). To conduct this experiment, the original density field taken from the n-body simulation has been used to compute an initial set of SFB coefficients (with $l_{max} = 1023$ and $N_{max} = 512$) and out to $r = 479 h^{-1}{\rm Mpc}$. Figure \ref{Figure:fieldOriginal} shows a slice of the 3D density field reconstructed from these coefficients. A Gaussian noise was then added to the SFB coefficients, the reconstruction of this noisy density field is shown on Figure \ref{Figure:fieldNoisy}. The hard thresholding algorithm was subsequently applied to the noisy SFB coefficients using 5 wavelet scales. The resulting density field is reconstructed on Figure \ref{Figure:fieldThresholded}. The residuals are shown on Figure \ref{Figure:residuals}. The wavelet analysis means we can successfully remove the noise we artificially added on entry , without much loss to the large scale structure, though some of the smaller structures are removed.

\section{Conclusion}
Modern cosmology requires the analysis of 3D fields on large areas of the sky, i.e. where the field is best viewed in spherical coordinates. In this configuration, a Spherical Fourier Bessel (SFB) transform is the most natural way to statistically analyse the field. Wavelet transforms have been shown to be ideally suited for cosmological fields, which tend to be sparse in wavelet space. The wavelet transform can be used e.g. for denoising, but there is yet no 3D wavelet transform in spherical coordinates.

We present in this paper a new 3D spherical wavelet transform, based on the undecimated wavelet transform (UWT) described in \citep{starck:sta05_2}. In order to perform operations on the wavelet transforms (such as denoising), we require a discrete version of the SFB transform for both the direct and inverse transforms. We show a novel way to perform such a fast Discrete Spherical Fourier-Bessel Transform (DSFBT) based on both a discrete Bessel transform and the HEALPIX angular pixelisation scheme. 

Using the 3D wavelet transform and the DSFBT, both introduced in this paper, we denoise a test large scale structure data set, taken from the Virgo large box simulations\footnote{\url{http://www.mpa-garching.mpg.de/Virgo/VLS.html}}. We find we can satisfactorily remove artificially added Gaussian noise without much loss to the large scale structure. All the algorithms presented in this paper are available for download as a public code called {\tt MRS3D} at \url{http://jstarck.free.fr/mrs3d.html}.

\begin{acknowledgements}
The authors are grateful to Boris Leistedt, Pirin Erdo\u gdu, Ofer Lahav and Alexandre R\'efr\'egier for useful discussions regarding the Spherical Fourier-Bessel decomposition and discretisation scheme. The 3D wavelet library uses Healpix software \citep{healpix:2002,gorski:2004by}. This research is in part supported by the Swiss National Science Foundation (SNSF).
\end{acknowledgements}

\bibliographystyle{aa}
\bibliography{JLSBibTex}

\appendix
\section{Spherical Fourier-Bessel Transform and 3D Convolution Products}
\subsection{Spherical Fourier-Bessel Transform and relation to the 3D Fourier Transform}\label{appendix:SFBT:Fourier}

In order to have a better understanding of the SFB coefficients and of how to use them to perform filtering, the SFB transform  can be related to the 3D Fourier transform. We follow a similar definition as the one presented in \cite{DSFBT}, but using our conventions for the different transforms.The following convention will be used for the Fourier transform:
\begin{equation}
F(\vec{k}) =\frac{1}{\sqrt{(2 \pi)^3}} \int f(\vec{r}) e^{-i \vec{k}.\vec{r}} d\vec{r} ,
\end{equation}
where $F$ denotes the Fourier transform of $f$. This formulation does not assume any coordinate system. However, to relate this transform to the SFB transform, it is possible to express this equation in spherical coordinates using the following expansion for the Fourier kernel:
\begin{equation}
e^{-i\vec{k}.\vec{r}} = 4\pi \sum_{l = 0}^{\infty} \sum_{m = -l}^{l} (-i)^l j_l (k r) \overline{ Y_l^m(\theta_r,\phi_r)} Y_l^m(\theta_k,\phi_k),
\end{equation}
where $(k, \theta_k,\phi_k )$ and $(r, \theta_r,\phi_r)$ are respectively the spherical coordinates of vectors $\vec{k}$ and $\vec{r}$. 

Substituting this expression for the kernel in the definition of the 3D Fourier transform yields: 
\begin{eqnarray}
F(k,\theta_k,\phi_k) & = & \frac{4\pi}{\sqrt{(2 \pi)^3}} \int_0^\infty \int_\Omega  \sum_{l = 0}^{\infty} \sum_{m = -l}^{l} (-i)^l f(r, \theta_r, \phi_r) \nonumber \\
    &   & \times \quad   j_l (k r)  \overline{Y_l^m(\theta_r,\phi_r)} Y_l^m(\theta_k,\phi_k) d\Omega r^2 dr \nonumber ,\\
	& = & \sum_{l = 0}^{\infty} \sum_{m = -l}^{l} (-i)^l \left[ \sqrt{\frac{2}{\pi}} \int_0^\infty \int_\Omega    f(r, \theta_r, \phi_r) \right. \nonumber \\
    &   & \times \quad \left.  j_l (k r)  \overline{Y_l^m(\theta_r,\phi_r)} d\Omega r^2 dr \right] Y_l^m(\theta_k,\phi_k) \nonumber ,\\
	& = & \sum_{l = 0}^{\infty} \sum_{m = -l}^{l}  \left[  (-i)^l \hat{f}_{l m} (k) \right] Y_l^m(\theta_k,\phi_k) .\label{Relationship_Fourier_Bessel}
\end{eqnarray}

In the last equation, the expression of the Spherical Harmonics Expansion of $F(k,\theta_k,\phi_k)$ for a given value of $k$ can be recognised from  Eq. (\ref{Inv_SHT}). In the Fourier space, the $(-i)^l \hat{f}_{l m}(k)$ are the Spherical Harmonics coefficients of $F$ on a sphere of given radius $k$. In other words, the Spherical Harmonics coefficients $F_{l m}(k)$ of the 3D Fourier transform $F(k, \theta_k,\phi_k)$ on a sphere of given radius $k$ in Fourier space are the SFB coefficients $\hat{f}_{l m}(k)$ for the same value $k$ but multiplied by factor $(-i)^l$.

The relationship between 3D Fourier transform and SFB transform is therefore very simple. The SFB transform can be sought of as a mere Fourier transform in spherical coordinates. In the next sections, this relationship will be used to derive convolution and filtering relations for the SFB transform using the well known relations verified by the Fourier transform.

\subsection{Expression of a 3D convolution product using the SFB Transform}\label{appendix:SFBT:convolution}

A prerequisite to the establishment of filtering relations is the expression of a 3D convolution product in terms of SFB coefficients. Let $v(r,\theta_r,\phi_r)$ be the 3D convolution of  $f(r,\theta_r,\phi_r)$ and $u(r,\theta_r,\phi_r)$. Then the 3D Fourier transform of $v$ verifies:
\begin{eqnarray}
V(k,\theta_k,\phi_k) & =  & \mathcal{F} \{ f \ast u \}(k,\theta_k,\phi_k)  \nonumber, \\
					 & =  & \sqrt{(2 \pi)^3 } F(k,\theta_k,\phi_k) U(k,\theta_k,\phi_k).
\end{eqnarray}
where $\mathcal{F}$ denotes the 3D Fourier transform. From Eq. (\ref{Relationship_Fourier_Bessel}) the expression of the 3D Fourier transform in spherical coordinates is known in terms of SFB coefficients. Applying this relationship to $V(k,\theta_k,\phi_k)$ in the last equation yields:

\begin{eqnarray}
	(-i)^l \hat{v}_{l m}(k) & = & \int_0^{2\pi} \int_0^{\pi} \sqrt{(2 \pi)^3 } F(k,\theta_k,\phi_k) U(k,\theta_k,\phi_k) \nonumber \\	
				  	&  & \times \quad \overline{Y_l^m (\theta_k, \phi_k)} \sin(\phi_k) d\phi_k d\theta_k  	.
\end{eqnarray}

Then, by applying Eq. (\ref{Relationship_Fourier_Bessel}) to $F$ and $U$ one gets:

\begin{eqnarray}
		\hat{v}_{l m}(k) 	 & = & (i)^l\sqrt{(2 \pi)^3 }\iint \sum\limits_{l'=0}^{\infty} \sum\limits_{m' = -l'}^{l'} (-i)^{l'} \hat{f}_{l' m'}(k) Y_{l'}^{m'}(\theta_k,\phi_k)\nonumber  \\
	& & \times \quad \sum\limits_{l''=0}^{\infty} \sum\limits_{m'' = -l''}^{l''} (-i)^{l''} \hat{u}_{l'' m''}(k) Y_{l''}^{m''}(\theta_k,\phi_k) \nonumber \\
 & & \times \quad  \overline{Y_l^m (\theta_k, \phi_k)} \sin(\phi_k) d\phi_k d\theta_k \nonumber ,\\
					 & = & (i)^l\sqrt{(2 \pi)^3 } \sum\limits_{l'=0}^{\infty} \sum\limits_{m' = -l'}^{l'} (-i)^{l'} \hat{f}_{l' m'}(k) \nonumber \\ 
					 & & \times \quad \sum\limits_{l''=0}^{\infty} \sum\limits_{m'' = -l''}^{l''} (-i)^{l''} \hat{u}_{l'' m''}(k) \nonumber \\
					 & &\times  \quad  \iint Y_{l'}^{m'} (\theta_k,\phi_k)  Y_{l''}^{m''}(\theta_k,\phi_k) \overline{Y_l^m (\theta_k, \phi_k)} d\Omega_k. 
\end{eqnarray}
The last integral over the two angular variables can be expressed as a Slater integral (which is a special case of the Gaunt integral) defined as:
\begin{equation}
	c^{l''}(l,m,l',m') = \iint \overline{Y_l^m(\theta,\phi)} 	Y_{l'}^{m'}(\theta,\phi) Y_{l''}^{m - m'}(\theta,\phi)  d\Omega.
\end{equation}
The Slater integrals are only nonzero for $| l - l' | \leq l'' \leq l + l'$ which simplifies the expression of $\hat{v}_{l m}(k)$.

The SFB transform of the 3D convolution product is therefore:
 \begin{eqnarray}
 	\widehat{(f \ast u)}_{l m}(k) & = & (i)^l \sqrt{(2 \pi)^3} \sum\limits_{l'=0}^{\infty} \sum\limits_{m' = -l'}^{l'} (-i)^{l'} \hat{f}_{l' m'}(k) \nonumber \\
 	& & \times \quad \sum\limits_{l''= | l - l' |}^{ l + l'} c^{l''}(l,m,l',m') (-i)^{l''} \hat{u}_{l'' m-m'}(k). \label{Spherical_Convolution}
 \end{eqnarray}

\end{document}